\documentclass[preprints,article,accept,pdftex,moreauthors]{Definitions/mdpi} 
\usepackage{subcaption}
\usepackage{subfig}
\usepackage[percent]{overpic}
\usepackage{longtable}
\usepackage{booktabs}
\firstpage{1} 
\pubvolume{1}
\issuenum{1}
\articlenumber{0}
\pubyear{2025}
\copyrightyear{2025}
\datereceived{ } 
\daterevised{ } 
\dateaccepted{ } 
\datepublished{ } 
\hreflink{https://doi.org/} 

\Title{Identifying Clean and Contaminated Atomic-Sized Gold Contacts under Ambient Conditions Using a Clustering Algorithm}

\TitleCitation{Title}

\Author{Guillem Pellicer $^{1}$\orcidB{}, Carlos Sabater $^{1,*}$\orcidA{}}

\isAPAStyle{%
       \AuthorCitation{Lastname, F., Lastname, F., \& Lastname, F.}
         }{%
        \isChicagoStyle{%
        \AuthorCitation{Lastname, Firstname, Firstname Lastname, and Firstname Lastname.}
        }{
        \AuthorCitation{Lastname, F.; Lastname, F.; Lastname, F.}
        }
}

\address{%
$^{1}$ \quad Departamento de F\'\i sica Aplicada and Instituto Universitario de Materiales de Alicante (IUMA), Universidad de Alicante, Campus de San Vicente del Raspeig, E-03690 Alicante, Spain.}

\corres{Correspondence: carlos.sabater@ua.es}

\abstract{Molecular electronics studies have advanced from early, simple single-molecule experiments at cryogenic temperatures to complex and multifunctional molecules under ambient conditions. However, room-temperature environments increase the risk of contamination, making it essential to identify and quantify clean and contaminated rupture traces (i.e., conductance versus relative electrode displacement) within large datasets. Given the high throughput of measurements, manual analysis becomes unfeasible. Clustering algorithms offer an effective solution by enabling automatic classification and quantification of contamination levels. Despite the rapid development of machine learning, its application in molecular electronics remains limited. In this work, we present a methodology based on the DBSCAN (Density-Based Spatial Clustering of Applications with Noise) algorithm to extract representative traces from both clean and contaminated regimes, providing a scalable and objective tool to evaluate environmental contamination in molecular junction experiments}


\begin{document}


\section{Introduction}
The core idea behind molecular electronics \cite{Agrait2003,Cuevasbook,Jan2019} is to use the smallest possible components---individual atoms or molecules---as active elements in electronic devices. A common strategy involves connecting a single atom or molecule between two electrodes. Break-Junction (BJ) techniques offer an excellent platform to achieve this level of control. The most widely used methods to measure electronic transport in such junctions are \textit{Scanning Tunneling Microscopy Break Junctions} (STM-BJ) \cite{Pascual1993} and \textit{Mechanically Controllable Break Junctions} (MCBJ) \cite{Krans93,SbRuitenbeek,Krans1995}. These techniques allow measurements electrical conductance $G$ the conductance (defined as $G = \frac{1}{R} = \frac{I}{V_{\text{bias}}}$).  Where $I$ is the current that follows the junctions and the $V_{bias}$ is the voltage applied to the junction.
According to Landauer’s formalism \cite{Landauer57}, the conductance of atomic and molecular junctions is quantized and can be expressed as $G = G_0 \sum_i T_i$, 
where \( T_i \) is the transmission probability of the \( i \)-th conduction channel, and \( G_0 = \frac{2e^2}{h} \) is the quantum of conductance \cite{Wees88} .   Here, \( e \) is the elementary charge, \( h \) is Planck’s constant, and the factor of 2 accounts for spin degeneracy.

In the case of single-atom metallic contacts---such as gold---the transmission probability is typically close to one, resulting in a conductance near $ G \approx G_0 \approx \frac{1}{12906} \, \Omega^{-1} \approx 7.75 \times 10^{-5} \, \text{S} $) \cite{AGrait93}. However, when the gold contact is further stretched, the junction breaks and the conductance abruptly drops to the tunneling regime, where the current decreases exponentially with the distance that separates the electrodes. When a molecule bridges the electrodes, the transmission is usually significantly reduced compared to metallic contacts (if the molecule is a poor conductor). The specific value depends on the molecular structure and its electronic coupling to the electrodes. In general, molecular junctions exhibit conductance values lower than $G_0$, often several orders of magnitude smaller \cite{Kamenetska22}.

The field of molecular electronics has undergone remarkable progress in recent decades~\cite{Agrait2003,Cuevasbook,Jan2019}, evolving from early demonstrations of single-molecule junctions at cryogenic temperatures~\cite{Smit2002,Elke11,Tewari19} to room-temperature molecular bridges with various functional properties~\cite{Tour97,Tao03,Herrer18,NicoMonte21,Ara24,Singh2025}. When measuring electronic transport in atomic-scale contacts under ambient conditions using BJ techniques, one of the main risks is sample degradation due to environmental exposure. Even brief contact with the laboratory atmosphere can lead to contamination of the junctions. Contamination that in form unknown molecule is captured between the leads when they are stretched or compressed. 

In parallel with advancements in molecular electronics, the past decade has seen a transformative rise in machine learning and artificial intelligence. While these tools have revolutionized numerous scientific fields, their application in molecular electronics started to grow in the last six years ~\cite{Cabosart19, Liu2020, Lin21, Solomon2022, Komoto23}. Our manuscript also is contributin to bring a new utilities for the molecular electronics using \textsc{DBSCAN}\cite{Ester96, Ester14, Schubert17}  clustering algorithm. This approach enables the automatic extraction of representative conductance-versus-displacement  that correspond to both clean and contaminated regimes in atomic-sized gold contacts measured under ambient conditions. By providing a robust statistical framework, our method allows for the distinction between pure metallic junctions and those compromised by environmental contamination, enhancing both the reproducibility and interpretability of room-temperature molecular electronics experiments.

Given the known challenges of obtaining ultra-clean atomic-scale junctions, one strategy to mitigate contamination is the \textit{in situ} cleaning of samples, for example through plasma treatment protocols. However, such cleaning techniques typically require disassembling the setup and removing the sample from the experimental chamber, which is not always feasible, especially during high-throughput or time-sensitive measurements.

In this article, we propose an alternative approach. Rather than physically cleaning the samples, we leverage a clustering algorithm to automatically classify conductance versus displacement traces---commonly referred to as conductance traces---according to whether they are clean (i.e., characteristic of pure metallic junctions) or contaminated (i.e., altered by molecular adsorption or ambient impurities). Our manuscript presents both the clustering protocol employed and its successful application to experimental data, clearly revealing the presence of two distinct classes of traces: those corresponding to clean gold junctions and those affected by contamination. This offers a powerful tool for assessing the cleanliness of the junctions during measurements, enabling the estimation of the proportion of clean traces in a dataset. In turn, this allows researchers to define thresholds for initiating sample cleaning or discarding data segments.

\section{Materials and Methods}
\subsection {Molecular electronics based on BJ experiments}
\begin{figure}[h]
\centering

\begin{overpic}[width=0.7\textwidth]{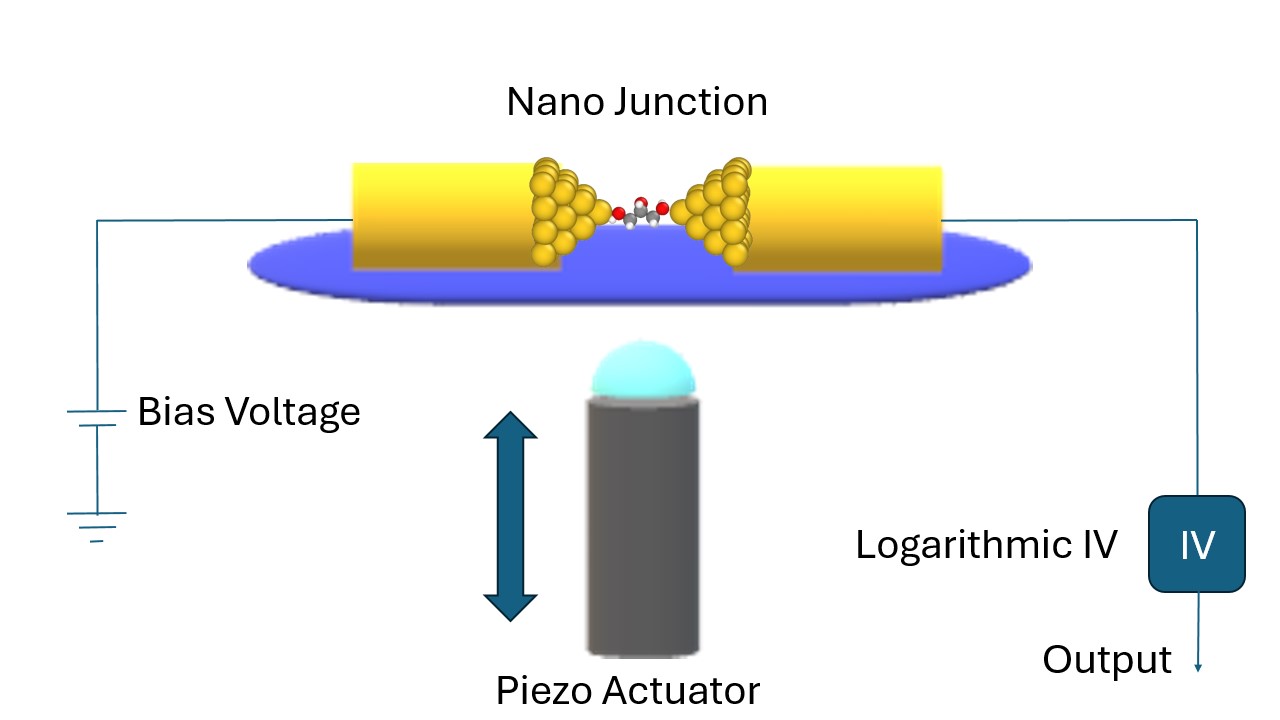}  \put(0,50){\textbf{(a)}}
\end{overpic}
\begin{overpic}[width=0.7\textwidth]{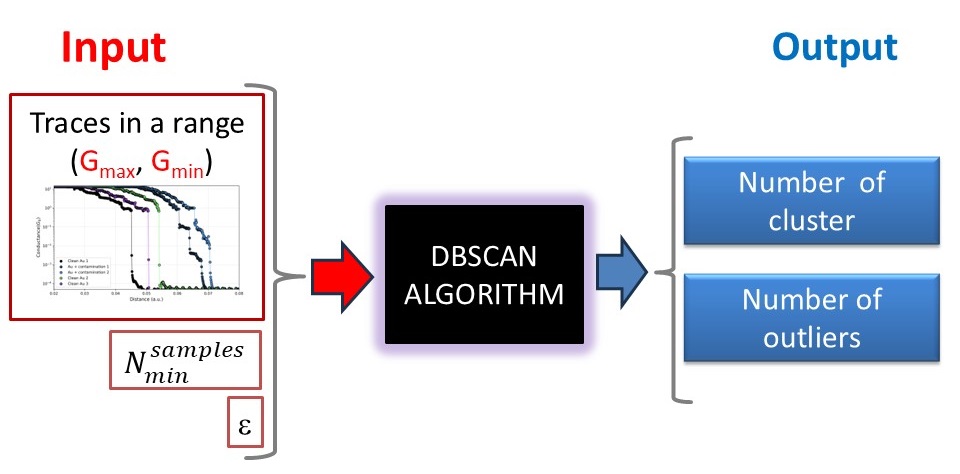} \put(0,50){\textbf{(b)}}
\end{overpic}
\caption{(\textbf{a}) Illustration and basic circuit of a MCBJ. 
(\textbf{b}) Input and output sketch of the DBSCAN algorithm.}
\label{setupRaw}
\end{figure}

To investigate atomic-scale electronic transport under ambient conditions, we employed a mechanically controllable break junction (MCBJ) setup (see Figure~\ref{setupRaw}a). This illustration shows a single molecule captured between two gold electrodes. A bias voltage is applied on the left side, and the current flows through a molecule trapped between the electrodes. These electrodes are mounted on a flexible polylactic acid (PLA) substrate. This substrate is bent using a piezoelectric actuator which, when pushed, causes the gold wire to break, and upon retraction, allows the flexible substrate to relax, bringing the electrodes back into contact and forming the junction again.

The current flowing through the molecule is extremely small, typically in the nanoampere or sub-nanoampere range. To accurately detect these low currents, a custom-built logarithmic I/V converter  was used \cite{Ornago2023}, allowing the amplification and recording of conductance signals as low as \(10^{-4} G_0\) . 

The system consists of a notched gold wire (Goodfellow, 0.1~mm diameter) carefully deposited and glued onto the flexible PLA substrate~\cite{Cuenca2025}. This configuration is integrated into a three-point bending mechanism, where a piezoelectric element provides precise mechanical control of the electrode separation, enabling the reproducible formation and rupture of atomic-scale junctions.

During the experiments, a constant bias voltage of 100~mV was applied across the junction. The resulting current was measured using the custom logarithmic I/V converter and recorded with a data acquisition (DAQ) system. This setup enables the acquisition of conductance traces, defined as the conductance versus relative electrode displacement—measured in volts or, when calibrated, in angstroms.

In this study, we focus exclusively on rupture traces. A total of approximately 5024 traces were acquired and analyzed under ambient conditions.

\subsection{Classification method: use of the DBSCAN algorithm}
Although there are various approaches to applying \textit{clustering} techniques to the classification of conductance traces, in this manuscript, we have chosen to use the DBSCAN algorithm. While the \textit{k-means} algorithm may be faster and more computationally efficient, we have preferred to sacrifice speed in favor of greater classification fidelity. DBSCAN is particularly effective when clusters exhibit relatively uniform density and there is a clear distinction between dense and sparse regions. Its operation is based on two key parameters: the neighborhood radius $\varepsilon$, which defines the maximum distance between two points to be considered neighbors, and the minimum number of samples $N_{\text{min}}^{\text{samples}}$ required for a point to be considered a core point.

In our case, the goal of clustering is not to classify individual points, but rather to identify and classify segments or ranges within each conductance trace. To this end, each segment is transformed into a feature vector, so the algorithm operates on these vectors rather than on the entire trace. This is illustrated in Fig.~\ref{setupRaw}(b), where the \textit{inputs} to the algorithm are the conductance ranges to be analyzed in each individual trace. It is important to note that, under this formulation, the parameters $\varepsilon$ and $N_{\text{min}}^{\text{samples}}$ are applied directly to the feature vector space, not to the individual points contained in the original traces.

As shown in Fig.~\ref{setupRaw}(b), the left side indicates the \textit{inputs}: the segmentation range, the number of variables considered, and the value of $N_{\text{min}}^{\text{samples}}$. Once a set of trace segments is processed (in our case, 5024 segments), and specific values of $\varepsilon$ and $N_{\text{min}}^{\text{samples}}$ are set, the algorithm returns the number of identified \textit{clusters}, as well as the number of traces that are not assigned to any cluster (classified as \textit{outliers}). One of the main advantages of DBSCAN is precisely its ability to automatically identify these \textit{outliers}.

Although DBSCAN does not require prior knowledge of the number of \textit{clusters}, having this information can be advantageous. In our case, we already know that the data ideally separates into two \textit{clusters}: one corresponding to clean traces and the other to contaminated ones. For a given segmentation range and specific parameters, we assess whether DBSCAN identifies these two groups. Any data  that does not fit well into either \textit{cluster} is classified as an \textit{outlier}. However, the optimal values of $\varepsilon$ and $N_{\text{min}}^{\text{samples}}$ are not known in advance.

Our goal is to distinguish between two types of conductance traces, or two \textit{clusters}: one corresponding to clean traces and the other to contaminated ones. As shown in Fig.~\ref{setupRaw}(b), the left panel illustrates the \textit{inputs}: the segmentation range, the number of variables considered, and the value of $N_{\text{min}}^{\text{samples}}$. Once a set of trace segments (in our case, 5024) is processed and specific values of $\varepsilon$ and $N_{\text{min}}^{\text{samples}}$ are set, DBSCAN returns the number of identified \textit{clusters} and the \textit{outliers}. In this work, we propose a systematic methodology to determine the optimal values of $\varepsilon$ and $N_{\text{min}}^{\text{samples}}$. To do so, we construct a meshgrid over the parameter space and run DBSCAN for all combinations of values. For each parameter pair, we record the number of \textit{clusters} and \textit{outliers}. To automate the process, we developed a \texttt{Python} script that explores the parameter space and generates a 3D map, where the $z$-axis indicates the number of \textit{clusters} and the color scale encodes the number of \textit{outliers}.  This visual tool provides an intuitive way to select an optimal parameter set for reliable classification. Once this plot is obtained, we refer to it as the result of a completed iteration.

However, in our protocol we must perform two iterations of DBSCAN in two distinct ranges to successfully classify the traces. This strategy has allowed us to develop a robust protocol that accurately distinguishes between ultraclean traces and those contaminated by the environment, even enabling estimation of the relative percentage of each type. It is worth noting that the complete application of the protocol takes no more than ten minutes per data set.


\section{Results}
\subsection {Data raw atomic-sized contacts of gold at room conditions}

\begin{figure}[H]
\centering
\begin{overpic}[width=0.45\textwidth]{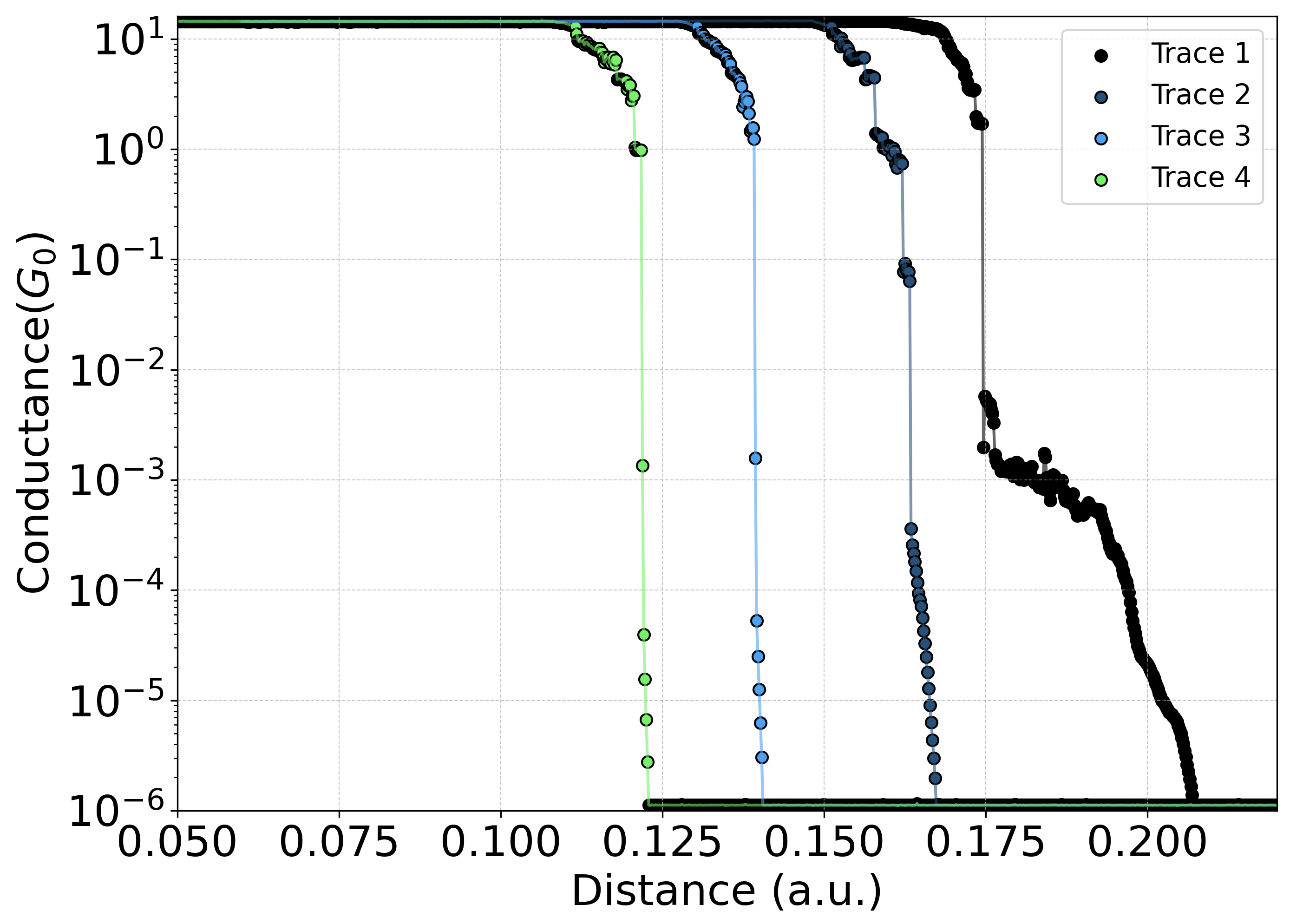}
  \put(0,75){\textbf{(a)}}
\end{overpic}
\hfill
\begin{overpic}[width=0.45\textwidth]{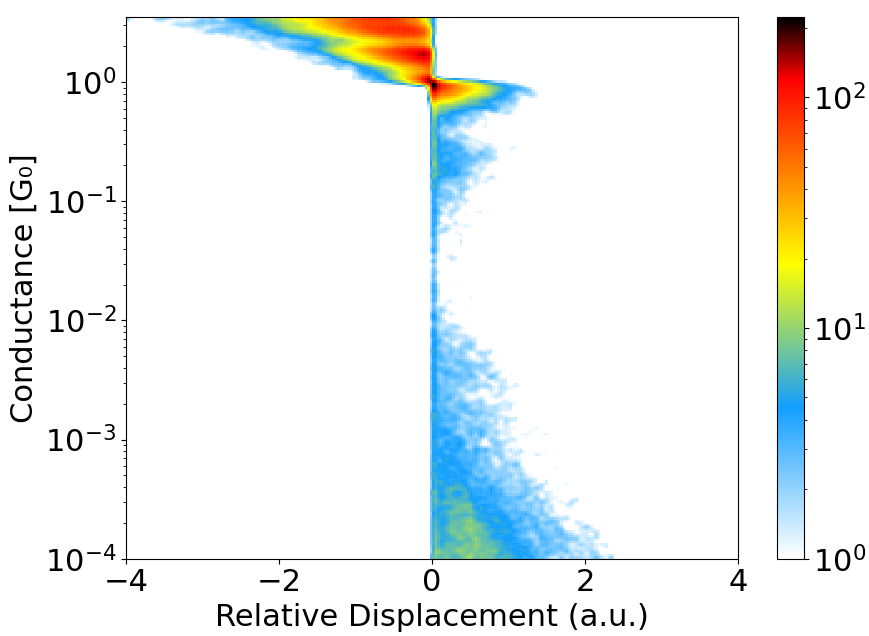}
  \put(0,75){\textbf{(b)}}
\end{overpic}
\caption{
\textbf{(a)} Representative conductance rupture traces plotted on a logarithmic scale. Traces 3 and 4 show direct transitions from \(1\,G_0\) to below \(10^{-4}\,G_0\), while traces 1 and 2 exhibit intermediate plateaus, indicating possible contamination. \textbf{(b)} Density map of 5024 gold rupture traces recorded at room temperature. The color scale reflects the number of events, with warm colors indicating high density and cool colors low density.
}
\label{RawData}
\end{figure}

Figure~\ref{RawData} (a) shows a selection of traces from the same experiment, plotted in logarithmic scale relative to displacement. It is well known that when an atomic contact is clean, the conductance drops sharply to the tunneling regime after the final atomic plateau at $1\,G_0$ upon further elongation \cite{Gimzewski87,J2CUntiedt, Sabater2013}. In contrast, when a molecule is present between the electrodes, a plateau often appears in the sub-quantum conductance range due to its low transmission \cite{Tao2003}.
The traces 1 and 2 of the figure .~\ref{RawData}(a) drop abruptly from \(1\,G_0\) to below \(10^{-4}\,G_0\), without exhibiting any intermediate plateaus; these ones can be classified as clean. In contrast, traces 3 and 4 display conductance plateaus between \(0.1\,G_0\) and \(10^{-4}\,G_0\), despite the absence of any intentionally deposited molecules. This behavior suggests the presence of contamination, potentially due to unintentional molecular junctions or residual species influencing the conductance.

An alternative statistical representation is the density color map, which can be constructed by aligning all traces with respect to the onset of the last plateau corresponding to the single-atom contact (in the case of gold \(1.0\,G_0\) ). Specifically, each trace is shifted such that the first point at \(1.0\,G_0\) is set to zero displacement. This procedure is applied to all traces, allowing them to share a common reference point.

Using this alignment method, a density map is generated, as shown in Fig.~\ref{RawData}(b), representing the full dataset composed of 5024 conductance traces. The map displays the number of events using a color scale: warm colors (dark red) indicate high occurrence density, while cool colors (blue) denote low density. The plot is constructed with 245 bins along the x-axis and 250 bins along the y-axis and is subsequently smoothed.

\subsection{Automated Selection of Optimal DBSCAN Parameters for Trace Classification}

\begin{figure}[H]
\centering
  \begin{overpic}[width=0.45\linewidth]{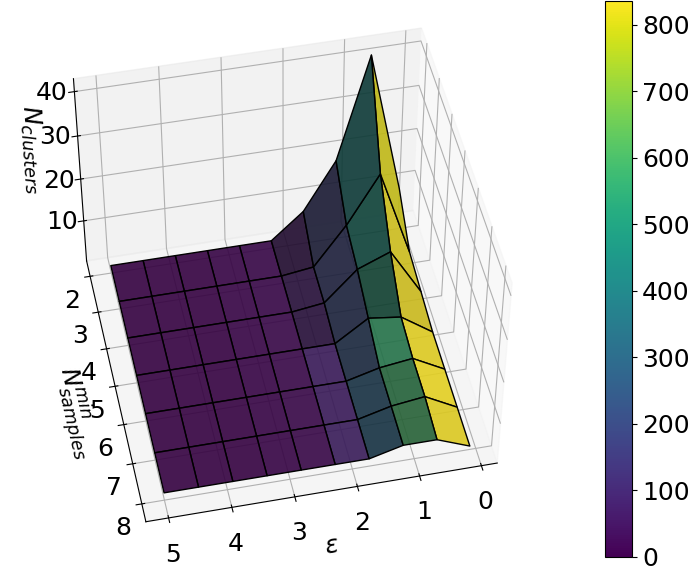}
    \put(0,75){\textbf{(a)}} 
  \end{overpic}
  \begin{overpic}[width=0.46\linewidth]{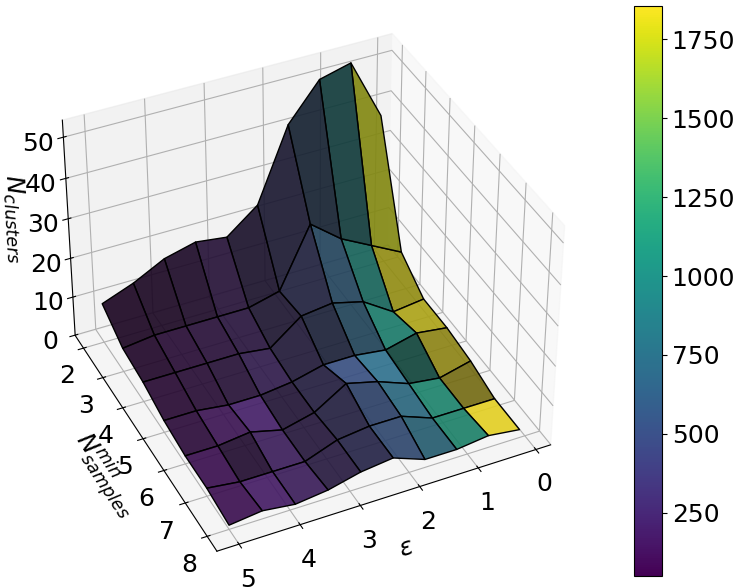}
    \put(0,75){\textbf{(b)}}
  \end{overpic}

\caption{Three-dimensional plots showing the number of clusters (\(z\)-axis) as a function of the DBSCAN parameters $\varepsilon$ and $N^{\text{min}}_{\text{samples}}$. The color scale indicates the number of outliers detected. \textbf{(a)} corresponds to the first iteration and \textbf{(b)} to the second iteration of DBSCAN.} 
\label{3Dplot}
\end{figure}

As explained earlier, DBSCAN depends on two parameters: \(\varepsilon\) and $N_{\text{min}}^{\text{samples}}$.  To choose them automatically, we created a Python script that scans the parameter space. For each pair of values, it runs DBSCAN and records the number of clusters and outliers. The results are shown in a 3D plot where the \(x\) and \(y\) axes represent the parameters, the \(z\)-axis shows the number of clusters, and the color scale indicates the number of outliers. This makes it easy to spot the most suitable region, as shown both panels Fig.~\ref{3Dplot}.

However, as previously noted in the \textit{Materials and Methods} section, a single DBSCAN iteration is not sufficient to accurately discriminate between fully clean and contaminated traces. For this reason, our protocol involves two successive DBSCAN runs, each applied to a different segment of the conductance traces.
In the first iteration, the conductance traces are segmented in the range $[10^{-1},10^{-3}]G_{0}$. This segment is used as input for a parameter sweep over $\varepsilon$ and $N^{\text{min}}_{\text{samples}}$ using a meshgrid approach. For each parameter pair, we compute the number of clusters and outliers identified by DBSCAN. The results are visualized in Fig.~\ref{3Dplot}(a).

From the 3D map, we identify an optimal region where the number of clusters is two and \( N^{\text{min}}_{\text{samples}} \) is between 4 and 7. Selecting parameters in this region helps the algorithm detect well-defined clusters associated with clean traces. Table~\ref{tab:parametrosDBSCAN1} shows a brief summary of the four selected possible DBSCAN parameter combinations explored.   The first column shows the number of the combination. Second and third columns show the values of $\varepsilon$ and $N^{\text{min}}_{\text{samples}}$, which yield exactly two clusters—matching the classification objective. The last three columns indicate the number of outliers, clean traces, and contaminated traces.

\begin{table}[H]
\centering
\caption{Summary of the possible DBSCAN parameter combinations explored in the \textbf{first} iteration}
\begin{tabularx}{\textwidth}{CCCCCCC}
\toprule
\textbf{Combination} & \textbf{$\varepsilon$} & \textbf{$N^{\text{min}}_{\text{samples}}$} & \textbf{Clusters} & \textbf{Outliers} & \textbf{Clean Traces} & \textbf{ Contaminated Traces} \\
\midrule
1 & 1.733 & 7 & 2 & 76 & 4188 & 760 \\
2 & 1.733 & 6 & 2 & 71 & 4188 & 765 \\
3 & 4.456 & 7 & 2 &  1 & 4188 & 835 \\
4 & 4.456 & 6 & 2 &  1 & 4188 & 835 \\
\bottomrule
\label{tab:parametrosDBSCAN1}
\end{tabularx}
\end{table}

As shown in Table~\ref{tab:parametrosDBSCAN1}, for \(\varepsilon\) values between approximately 1.5 and 4.5 and \(N^{\text{min}}_{\text{samples}}\) between 6 and 7, the number of traces classified as clean remains constant at 4188. Within this range, the number of \textit{outliers} varies between 1 and 76, while the number of contaminated traces ranges from 760 to 836, always over a total of 5024 traces. In this first iteration, we selected combinations 3 and 4, as both yield the minimum number of \textit{outliers} (1) and allow the remaining traces to be classified as contaminated without affecting the number of clean traces. Since both combinations produce equivalent results, we adopted option 3 (or 4) as the reference for the next iteration. However, as clearly shown in Fig.~\ref{densplots}(a), a noticeable cloud of counts remains in the region around $10^{-1}G_{0}$, indicating that the separation is not entirely clean. In fact, the data raw density plot in Fig.~\ref{RawData}(b), which includes all unfiltered traces (including contaminated ones), appears strikingly similar to the distribution obtained after the first DBSCAN filtering shown in Fig.~\ref{densplots}(a). Therefore, a second iteration is warranted, as previously outlined in the Materials and Methods section of this manuscript.

In the second iteration, the same procedure is applied to a different segment of the conductance traces, $[10^{0},10^{-1}]G_{0}$, over the 4188 clean traces. The results obtained are shown in Fig.~\ref{3Dplot}(b). This second analysis serves to refine the classification and helps distinguish really clean traces or traces with contaminants, completing the overall protocol.

As in the first iteration, we compile a summary table to evaluate different DBSCAN parameter combinations that result in two clusters. Table~\ref{tab:parametrosDBSCAN2} presents these combinations. Based on this table, we observe that for very small values of \(\varepsilon\), the number of outliers increases significantly. Upon inspection, this behavior appears to misclassify some genuinely clean traces as outliers. On the other hand, when \(\varepsilon\) is too large, the number of outliers drops drastically. However, visual inspection of the supposedly clean traces reveals the presence of residual sub-plateaus in certain regions, suggesting that some contaminated traces are being misclassified as clean. Therefore, we select an intermediate value of \(\varepsilon = 3.5\), which results in a reasonable number of outliers and a sufficiently high number of clean traces. Specifically, out of the initial 4188 clean traces, 3912 are retained after this second filtering step.

\begin{table}[H]
\caption{Summary of the possible DBSCAN parameter combinations explored in the \textbf{second} iteration}
\begin{tabularx}{\textwidth}{CCCCCCC}
\toprule
\textbf{Combination} & \textbf{$\varepsilon$} & \textbf{$N^{\text{min}}_{\text{samples}}$} & \textbf{Clusters} & \textbf{Outliers} & \textbf{Clean Traces} & \textbf{ Contaminated Traces} \\
\midrule
1 & 0.644 & 6 & 2 & 1103 & 3077 & 6 \\
2 & 3.400 & 7 & 2 & 218 & 3912 & 56 \\
3 & 5.000 & 6 & 2 & 91 & 4087 & 8 \\

\bottomrule
\label{tab:parametrosDBSCAN2}
\end{tabularx}
\end{table}

\subsection{Clustering Based on DBSCAN to Identify Pure Metallic Atomic-Scale Gold Contacts and Traces with Contamination}

\begin{figure}[H]
\centering
\begin{overpic}[width=0.45\linewidth]{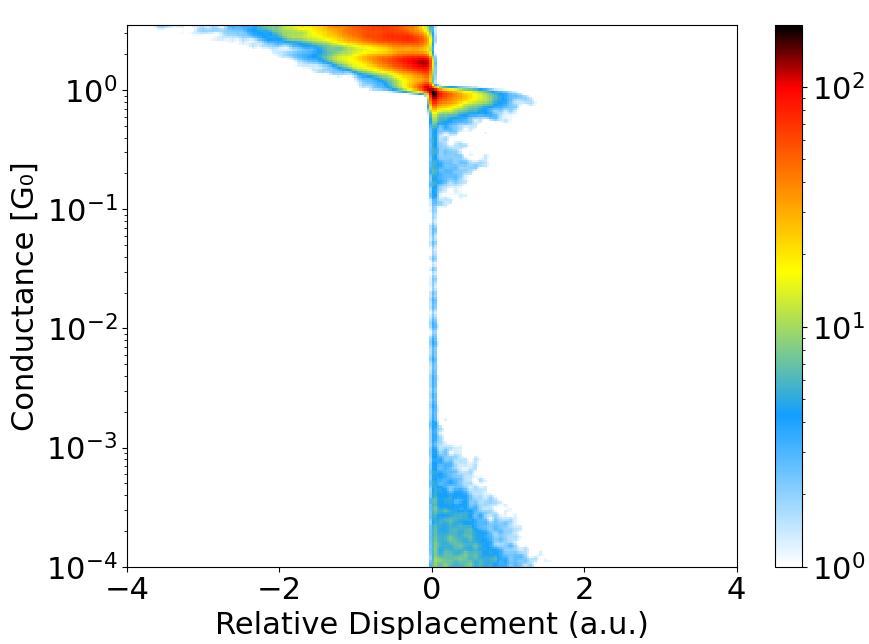}
\put(0,75){\textbf{(a)}} 
  \end{overpic}
\begin{overpic}[width=0.45\linewidth]{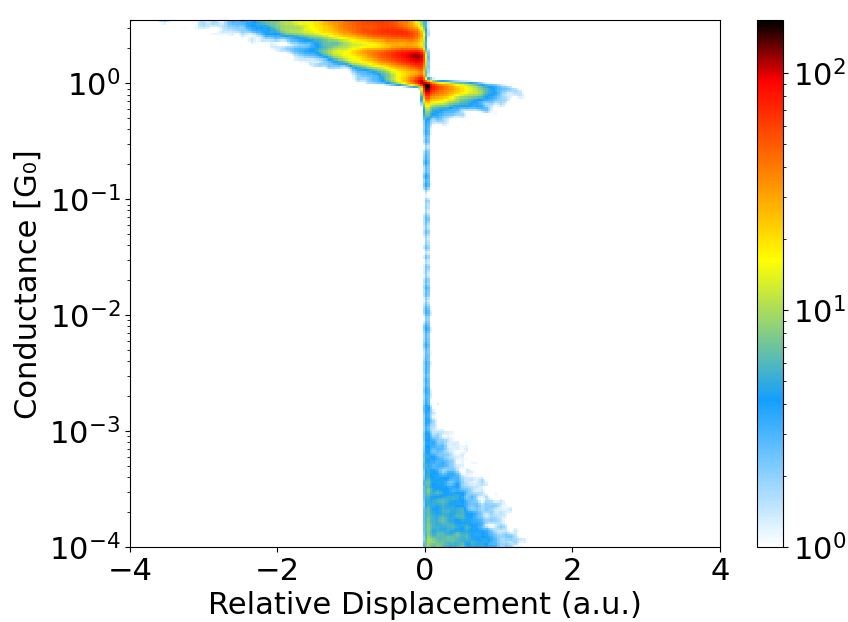}
   \put(0,75){\textbf{(b)}} 
  \end{overpic}
\caption{Panel (\textbf{a}) shows a density plot constructed from the 4188 clean traces identified in the first iteration using combination 3–4 from Table~\ref{tab:parametrosDBSCAN1}. Panel (\textbf{b}) displays a density plot based on the 3912 clean traces identified in the second iteration using combination 2 from Table~\ref{tab:parametrosDBSCAN2}. The number of bins used for the density plots are 250 and 300, respectively.}
\label{densplots}
\end{figure}

Once the DBSCAN algorithm has been applied to classify the traces as clean or contaminated, one of the most effective ways to evaluate its performance is to represent both iterations using density plots. Figure~\ref{densplots} displays the results: panel~(a) shows the density plot from the first iteration, while panel~(b) corresponds to the second iteration, in which traces identified as completely clean are shown.

As can be observed, the contaminated traces shown in Fig.~\ref{densplots} panel~(a) behave quite differently. Around \(10^{-1}\,G_0\), a bluish cloud is clearly visible, indicating contamination. In contrast, the density plot in panel~(b) exhibits very few counts within a noticeable gap between the atomic contact of gold (\(\sim 1\,G_0\)) and the tunneling regime (which starts around \(10^{-3}\,G_0\) and is characterized by its slope on the logarithmic scale). Although a small number of points remain in this clear gap, the number of counts in this gap is three orders of magnitude compared to the maximum. These isolated points are typically attributed to artifacts from the data acquisition system (DAQ).

These qualitative differences between the two density plots confirm that they reflect distinct physical behaviors. A more detailed discussion supporting the validity of our classification algorithm is provided in the following section.

\section{Discussion}
Thanks to the DBSCAN clustering algorithm applied following the protocol described above, we have been able to classify conductance traces into two well-defined groups: those confidently labeled as \textit{clean}—even under ambient conditions—and those interpreted as affected by \textit{contamination}. This classification is clearly reflected in the density plots shown in Figure~\ref{densplots}. Panel~(a) corresponds to the first iteration, which fails to correctly identify clean traces, while panel~(b) shows the result of the second iteration, where clean traces are clearly isolated.

\begin{figure}[H]
\centering
\begin{overpic}[width=0.43\linewidth]{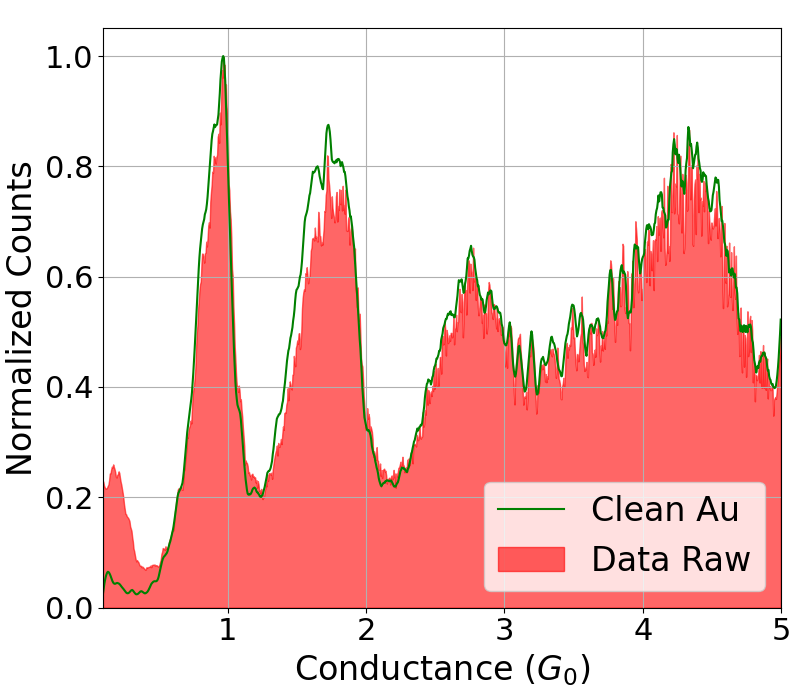}
\put(0,84){\textbf{(a)}} 
  \end{overpic}
\begin{overpic}[width=0.45\linewidth]{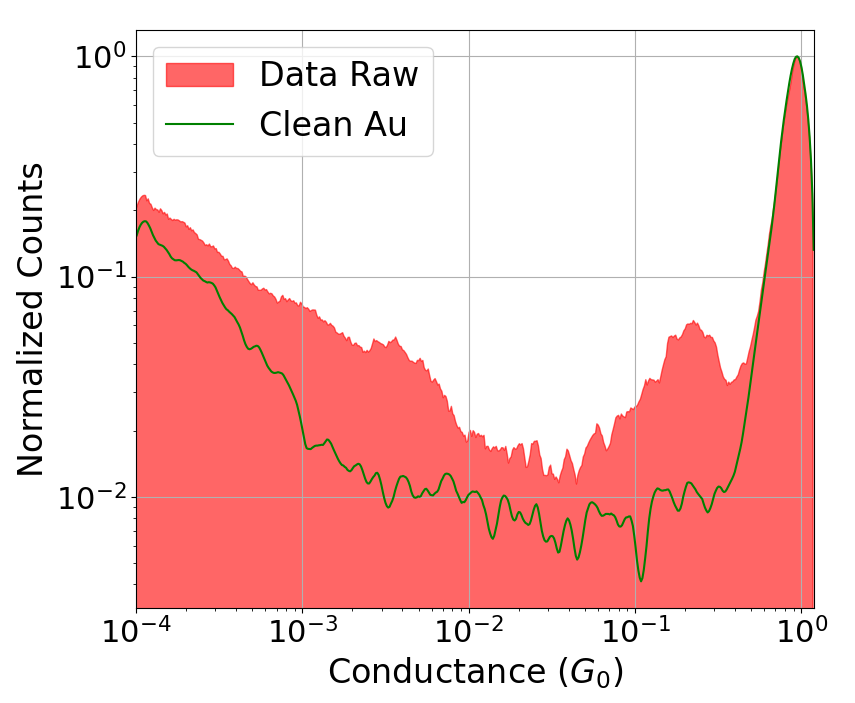} \put(0,82){\textbf{(b)}} 
  \end{overpic}
\caption{Panel (a) shows the histogram on a linear scale, while panel (b) presents it on a logarithmic scale (log-log). The red curves correspond to the raw dataset (5024 traces), and the green curves represent the subset of clean 3912 traces identified by the DBSCAN algorithm.}
\label{HistosLinLog}
\end{figure}

Moreover, we want to bring clarity to our message and to be more conclusively demonstrate the cleanliness of the selected traces. Therefore, we propose an additional statistical analysis based on normalized conductance histograms (to one quantum of conductance), presented in both linear and logarithmic scales (see Fig. \ref{HistosLinLog}). In these histograms, we compare the raw dataset with the subset of 3912 traces identified as clean. The color code indicates red for the data raw and green for the classified as clean.

The linear-scale histogram allows us to observe high conductance features and directly compare the overall profile of the raw(red) and clean datasets (green). In contrast, the log-log histogram (logarithmic in both counts and conductance) enhances the visibility of features typically associated with environmental contamination, such as peaks around \(  \sim 1 \cdot 10^{-1}\,G_0\) or even down to \( \sim 1 \cdot 10^{-4}\,G_0\). As expected, the log-log representation emphasizes all differences, especially in the low-conductance region. Finally, we provide a comparative table showing the number of traces classified as clean and contaminated within our dataset.

Further assess the robustness of our clustering procedure, we analyze the conductance histograms in the range from 0.1 to 5~$G_0$ for both the raw data(red color) and the clean subset (green line). 
As shown in Figure~\ref{HistosLinLog}(a), the histogram of the full dataset  exhibits a strong peak centered around 1~$G_0$, consistent with the typical conductance of single-atom Au contacts. 
Focusing on the peaks at 1, 2, and 3~$G_0$, we observe that, for both the raw data and the clean dataset, the peak positions remain essentially the same, with only subtle differences. A slight decrease in the count density is noticeable near 2~$G_0$ for the clean traces. Interestingly, and in contrast with previous studies where molecules are deposited on gold atomic contacts—where a significant broadening of the 1~$G_0$ peak is typically observed  \cite{Ara22,Martinez23}—such broadening is absent in our data, suggesting that environmental contamination does not produce a comparable effect.

However, a notable difference emerges in the low-conductance regime, around 0.1~$G_0$. In this region, a clear peak appears in the raw dataset but is entirely absent in the clean traces. This highlights the importance of representing conductance histograms in a logarithmic scale, which allows better resolution in the low-conductance region and helps reveal the true positions and nature of such peaks, potentially masked in linear representations.

The distinction between clean and contaminated traces is clearly observed in the log-log scale histograms in Fig. ~\ref{HistosLinLog}(b). In the clean dataset (represented by the green histogram), there's a marked deep in the count density, specifically between approximately $8 \cdot 10^{-1}~G_0$ and $1 \cdot 10^{-3}~G_0$. Below this threshold, the histogram exhibits a linear trend on the log-log scale, which is a characteristic behavior of the tunneling regime. Conversely, the raw data (red area) displays a prominent distribution of conductance values within the same range. Here, a significant concentration of events stands out, centered around $2 \cdot 10^{-1}~G_0$. This peak is commonly associated with contamination from environmental sources or hydrocarbons, as confirmed in our previous study~\cite{Ara22}. It's crucial to note that the count density of this peak in the contaminated histogram, close to $2 \cdot 10^{-1}~G_0$, is almost an order of magnitude higher compared to the clean traces. The latter, in contrast, show no distinct peak in this region, only baseline counts. Furthermore, another prominence  exist in the contaminated data also disrupts the linear trends observed in the clean traces within the $10^{-2}~G_0$ to $10^{-3}~G_0$ range. This deviation is also attributed to contamination, given that no molecules were intentionally deposited and, under ideal conditions, no counts should appear in this range.

With all this data, we can quantitatively determine the percentage of clean and contaminated traces obtained after applying our classification process. This statistical analysis allows us to evaluate whether our samples progressively degrade over time due to environmental exposure, or if contamination occurs immediately upon exposure.

The following table \ref{tab:traceStats} summarizes the total number of traces analyzed, along with the relative proportions of clean and contaminated traces:

\begin{table}[H]
\centering
\caption{Statistical summary of trace classification}
\begin{tabularx}{\textwidth}{CCC}
\toprule
\textbf{Total traces} & \textbf{Clean traces (\%)} & \textbf{Contaminated traces (\%)} \\
\midrule
5024 & 3912 (77.8\%) & 1112 (22.2\%) \\
\bottomrule
\label{tab:traceStats}
\end{tabularx}
\end{table}

Altogether, the combined visual and statistical analyses support the effectiveness of our DBSCAN-based classification. The subset of clean traces not only excludes contamination artifacts but also reproduces features known from high-purity, low-temperature measurements. This reinforces the reliability of our approach and demonstrates that meaningful structural and electronic information can be extracted from ambient-condition experiments, provided that appropriate data-cleaning procedures are applied.

\section{Conclusions}
Therefore, we conclude that we have developed a robust trace classification procedure capable of separating clean and contaminated traces through a two-step DBSCAN-based protocol. This methodology enables a quantitative evaluation of the proportion of clean versus contaminated traces within a dataset. In our case, we determined that $77.8\%$ of the traces are completely clean, meaning they do not exhibit any counts or plateaus in the range of [$8 \cdot 10^{-1}, 1 \cdot 10^{-3}$] $G_{0}$.

Importantly, this constitutes the first compact device that, via electronic transport measurements, can quantify the presence of a single contaminant molecule. Although it does not provide information about the chemical identity of the contaminant, it reliably assesses its abundance.

These findings demonstrate that a large fraction of the traces are of high purity. Furthermore, our methodology could be extended to evaluate how different metals degrade upon exposure to environmental conditions. This, in turn, would support decisions about whether molecular deposition is justified or help define the minimum contamination threshold required for reproducible and reliable molecular-scale measurements.



\vspace{6pt} 





\authorcontributions{ G.P. contributed to the methodology, software development, investigation, and data analysis. C.S. contributed to the conceptualization, methodology, validation, formal analysis, investigation, resources, supervision, writing—original draft, writing—review and editing, and funding acquisition. }

\funding{This research received no external funding}

\acknowledgments{This work received financial support from the Generalitat Valenciana through CIDEXG/2022/45 and the Spanish Government by PID2023-146660OB-I00. This research is an integral part of the Advanced Materials program, supported by MCIN with funding from the European Union NextGenerationEU (PRTR-C17.I1) and the Generalitat Valenciana (MFA/2022/045). We also acknowledge funding from MICIU/AEI/10.13039/501100011033 and the European Regional Development Fund (ERDF/EU) under project PID2023-146660OB-I00. Finally, the authors extend their gratitude for insightful discussions with Prof. Carlos Untiedt, Dr. Wynand Denam, Dr. Tamara de Ara, Andrés Martínez, Juan Pablo Cuenca, and Enrique Guzmán.}

\conflictsofinterest{The authors declare no conflicts of interest. } 



\abbreviations{Abbreviations}{
The following abbreviations are used in this manuscript:
\\

\noindent 
\begin{tabular}{@{}ll}
MCBJ & Mechanical Controllable Break Junctions\\
DBSCAN & stands for Density-Based Spatial Clustering of Applications with Noise
\end{tabular}
}

\isPreprints{}{
\begin{adjustwidth}{-\extralength}{0cm}
} 

\reftitle{References}


 \bibliography{Refclean.bib}

\isPreprints{}{
\end{adjustwidth}
} 

\end{document}